\documentclass[twocolumn,showpacs,preprintnumbers,superscriptaddress]{revtex4}
\usepackage{graphicx,dcolumn,bm,amsmath,amssymb}

\newcommand{\Sec}[1]{Sec.\,\ref{#1}}

\newcommand{\fref}[1]{Fig.\,\ref{#1}}
\newcommand{\Fref}[1]{Fig.\,\ref{#1}}
\newcommand{\eref}[1]{Eq.\,(\ref{#1})}
\newcommand{\nl}{\nonumber \\}
\newcommand{\be}{\begin{equation}}
\newcommand{\ee}{\end{equation}}
\newcommand{\bea}{\begin{eqnarray}}
\newcommand{\eea}{\end{eqnarray}}
\newcommand{\bsube}{\begin{subequations}}
\newcommand{\esube}{\end{subequations}}
\newcommand{\alf}{\alpha}
\newcommand{\sgm}{\sigma}
\newcommand{\omg}{\omega}
\newcommand{\Omg}{\Omega}
\newcommand{\Gam}{\Gamma}
\newcommand{\vpl}{\varepsilon}
\newcommand{\upa}{\uparrow}
\newcommand{\dwa}{\downarrow}
\newcommand{\la}{\langle}
\newcommand{\ra}{\rangle}
\newcommand{\rmL}{{\rm L}}
\newcommand{\rmR}{{\rm R}}
\newcommand{\rmc}{{\rm c}}
\newcommand{\rmd}{{\rm d}}
\newcommand{\rmi}{{\rm i}}
\newcommand{\Tr}{{\rm Tr}}
\newcommand{\GamL}{\Gamma_{\rm L}}
\newcommand{\GamR}{\Gamma_{\rm R}}

\begin{document}

 \title{Spin-resolved bunching and noise characteristics in double quantum
 dots coupled to ferromagnetic electrodes}

 \author{JunYan Luo}
 \email{jyluo@zust.edu.cn}
 \affiliation{School of Science, Zhejiang University of Science
  and Technology, Hangzhou 310023, China}

 \author{HuJun Jiao}
 \affiliation{Department of Physics, Shanxi University, Taiyuan, Shanxi 030006, China}

 \author{BiTao Xiong}
 \affiliation{School of Science, Zhejiang University of Science
  and Technology, Hangzhou 310023, China}

 \author{Xiao-Ling He}
 \affiliation{School of Science, Zhejiang University of Science
  and Technology, Hangzhou 310023, China}

 \author{Changrong Wang}
 \affiliation{School of Science, Zhejiang University of Science
 and Technology, Hangzhou 310023, China}

\date{\today}

\begin{abstract}
 We study spin-resolved noise in Coulomb blockaded double quantum
 dots coupled to ferromagnetic electrodes.
 The modulation of the interdot coupling and spin polarization
 in the electrodes gives rise to an intriguing dynamical spin
 $\upa$-$\upa$ ($\dwa$-$\dwa$) blockade mechanism: Bunching of
 up (down) spins due to dynamical blockade of an up (down) spin.
 In contrast to the conventional dynamical spin $\upa$-$\dwa$ bunching
 (bunching of up spins entailed by dynamical blockade of a down spin),
 this new bunching behavior is found to be intimately
 associated with the spin mutual-correlation, i.e., the noise
 fluctuation between opposite spin currents.
 We further demonstrate that the dynamical spin $\upa$-$\upa$ and
 $\upa$-$\dwa$ bunching of tunneling events may be coexistent in the regime of weak
 interdot coupling and low spin polarization.
 \end{abstract}



\maketitle

\section{\label{thsec1}Introduction}

 The measurement of signal-to-noise ratio in mesoscopic
 transport devices is of vital importance as it enables
 access to intriguing information about the statistics of
 quasiparticles and various intrinsic dynamics that are
 not available from conventional current measurements
 alone \cite{Bla001,Naz03}.
 For transport through a localized state, the nonequilibrium
 noise is generally suppressed due to the
 Pauli exclusion principle, leading thus to a
 sub-Poissonian statistics \cite{Che914534,Oli99299,Hen99296}.
 However, in systems of multiple nonlocal states, such
 as double quantum dot devices, the
 intrinsic quantum coherence and many-particle interactions
 give rise to different sources of correlations \cite{Wie031}.
 Electron transport can exhibit a unique dynamical
 blockade mechanism, leading thus to a super-Poisson
 fluctuation \cite{Kie07206602,Lin09245303,Mic09035320,%
 Luo11145301,Luo1159,Lam08214302}.

 In spintronic structures, transport is governed not
 only by the charge flow, but more importantly, by the
 spin transfer \cite{Pri981660,Wol011488,Jed01345,%
 Jed02713,Aws02,Mor11345,Han071217,Zut04323}.
 The involving spin degrees of freedom introduce
 additional correlated mechanisms.
 Study of spin current fluctuations is crucial for
 possible applications in control and manipulation
 of individual spins.
 A number of investigations have been devoted to the
 noise characteristics of spin-dependent transport through
 various nanostructures, such as quantum
 dots \cite{Han071217,Zut04323,Thi05146806,Bra06075328},
 molecules \cite{Wey12205306,Yu05075351,Mis09224420}, and
 nanotubes \cite{Wey10165450,Wey08035422,Lip10115327,Wu07156803}.
 Different tunneling processes like sequential tunneling,
 cotunneling \cite{Wey11195302,Wey08075305}, etc. were revealed
 to have vital roles to play in spin transport.
 In order to distinguish various spin dynamics it is
 instructive to unravel the charge noise into
 spin-resolved components.
 Consider a general mesoscopic device of a quantum
 dot (QD) system connected to terminals $\alf$, $\alf'$....
 The charge current through the terminal ``$\alf$'' is
 $\la I_\alf\ra=\la I_\alf^{\sgm}\ra+\la I_\alf^{-\sgm}\ra$,
 where $\la I_\alf^{\sgm}\ra$ is the spin-$\sgm$ component
 of the current.
 The temporal correlation of spin transport is characterized
 by the spin-resolved correlation function
 $C^{\sgm\sgm'}_{\alf\alf'}(t-t')=\frac{1}{2}
 \la\{\Delta I_\alf^\sgm(t),\Delta I_{\alf'}^{\sgm'}(t')\}\ra$,
 with $\Delta I_\alf^\sgm(t)=I_\alf^\sgm(t)-\la I_\alf^\sgm(t)\ra$.
 Straightforwardly, the individual spin-resolved noise power
 is given by
 $S_{\alf\alf'}^{\sgm\sgm'}(\omg)=\int_{-\infty}^{\infty}\rmd t\,
 e^{\rmi \omg t}C^{\sgm\sgm'}_{\alf\alf'}(t)$.
 By choosing an arbitrary spin axis \textbf{z}, the total
 charge current noise can be readily unraveled into
 $S_{\alf\alf'}^{\rm ch}=
 S_{\alf\alf'}^{\upa\upa}+S_{\alf\alf'}^{\dwa\dwa}+
 S_{\alf\alf'}^{\upa\dwa}+S_{\alf\alf'}^{\dwa\upa}$.
 Here, the spin self-correlation $S_{\alf\alf'}^{\upa\upa}$
 ($S_{\alf\alf'}^{\dwa\dwa}$) represents fluctuation
 between the same spin currents, which was recently shown
 to be capable of serving as a sensitive tool to identify
 the dynamical spin $\upa$-$\dwa$
 ($\dwa$-$\upa$) bunching, i.e., bunching of up (down) spins
 due to dynamical blockade of a down (up)
 spin \cite{Cot04115315,Cot04206801,Don09153305}.
 It then naturally comes to a question:
 What can we infer from the spin
 mutual-correlation noise (fluctuation between opposite
 spin currents) $S_{\alf\alf'}^{\upa\dwa}$ or
 $S_{\alf\alf'}^{\dwa\upa}$?

 \begin{figure}
 \begin{center}
 \includegraphics*[scale=0.7]{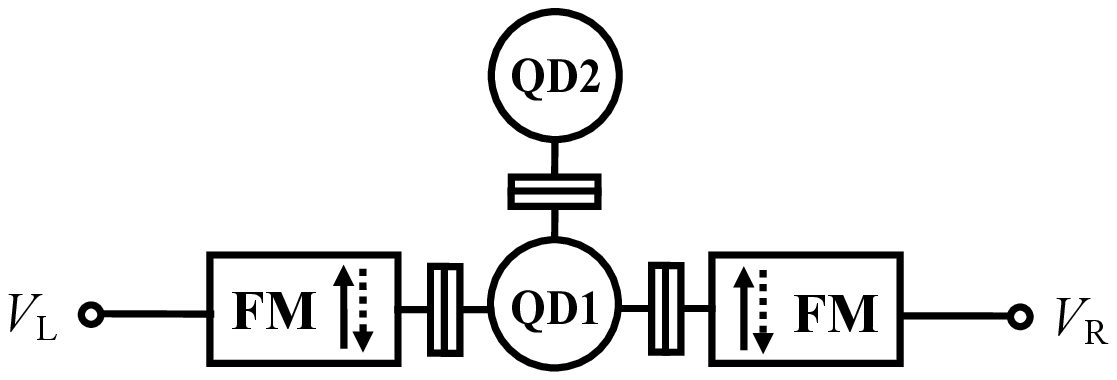}
 \caption{\label{Fig1}Schematic setup for
 spin-dependent transport through double
 quantum dots, in which only QD1 is tunnel-coupled
 to ferromagnetic electrodes.}
 \end{center}
 \end{figure}

 \begin{figure*}
 \begin{center}
 \includegraphics*[scale=1]{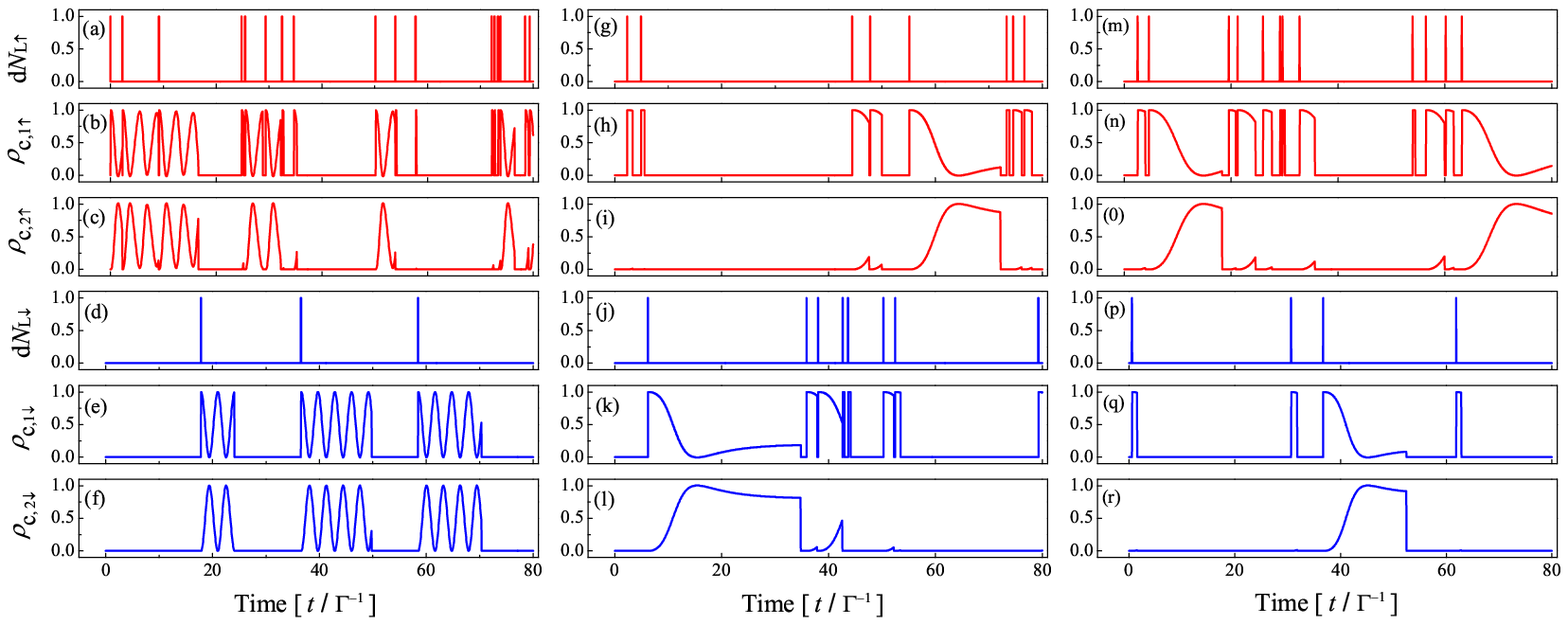}
 \caption{\label{Fig2}
 Sets of typical detection record (up and down spin tunneling
 events) and corresponding real-time quantum states under
 parallel magnetic alignment.
 (a)-(f) $\Omg/\Gam=1.0$ and $p=0.8$,
 (g)-(l) $\Omg/\Gam=0.2$ and $p=0$, and
 (m)-(r) $\Omg/\Gam=0.1$ and $p=0.25$.
 Each quantum dot has only one level (assumed to be in resonance,
 i.e., $E_1=E_2$) within the bias window $V=V_\rmL-V_\rmR$.
 In the Coulomb blockade regime and for temperature $k_{\rm B} T\ll V$,
 the Fermi functions can be approximated by either one or
 zero, so the temperature is not involved here.}
 \end{center}
 \end{figure*}

 In the context of spin-dependent transport through a system of multiple
 nonlocal states, we investigate in this work these
 spin-resolved noise components and their connections to
 spin-resolved bunching behavior.
 Specifically, we consider a double quantum dot,
 where only one
 of the dots is tunnel-coupled to the ferromagnetic (FM)
 electrodes (see \fref{Fig1}).
 The system, which can be mapped to the one investigated
 recently in experiments \cite{Nau02161303,Saf03136801,%
 Nau04113316,Sas09266806}, is of particular interest, as
 it is arranged in such a configuration that can maximize
 locality versus nonlocality
 contrast \cite{Kim01245326,Cor05075305,Dju05032105,Luo10083720}.
 In the Coulomb blockade regime, we observe a unique
 dynamical spin $\upa$-$\upa$ ($\dwa$-$\dwa$) blockade
 phenomenon, namely, bunching of up (down) spins due to
 dynamical blockade of an up (down) spin.
 Different from the conventional spin $\upa$-$\dwa$ bunching,
 it is revealed that this new mechanism is intimately
 associated with positive spin mutual-correlation.
 We further demonstrate that the spin $\upa$-$\upa$ and
 $\upa$-$\dwa$ bunching of tunneling events may be
 coexistent in the regime of low spin polarization and
 weak interdot tunnel-coupling.

 The paper is organized as follows.
 In \Sec{thsec2}, we describe the double quantum dot
 system tunnel-coupled to FM electrodes.
 In \Sec{thsec3}, a Monte Carlo approach is introduced
 to simulate real-time single spin tunneling events.
 The spin-resolved noises, together with spin-resolved
 bunching of tunneling events will be discussed
 in detail.
 It is then followed by the conclude in \Sec{thsec4}.

 \section{\label{thsec2}Model Description}

 The system under study is sketched in \fref{Fig1},
 in which only QD1 is connected to the FM electrodes, whereas
 QD2 is side-connected to the QD1.
 The Hamiltonian of the entire system is
 $\hat{H}=\hat{H}_{\rm B}+\hat{H}_{\rm S}+\hat{H}'$,
 where $\hat{H}_{\rm B}=\sum_{\alf={\rm L,R}}\sum_{k,\sgm}
 \vpl_{\alf k\sgm}c_{\alf k\sgm}^\dag c_{\alf k\sgm}$ denotes 
 the left and right FM electrodes;
 $c_{\alf k\sgm}$ ($c_{\alf k\sgm}^\dag$) is the
 electron annihilation (creation) operator of the electrode
 $\alf=$ L or R, with spin $\sgm=\,\upa$ or $\dwa$.
 The ferromagnetism of the electrode $\alf$ is
 accounted for by the spin-dependent density of
 states $g_{\alf\sgm}(\omg)$.
 Throughout all of our calculations presented here,
 we approximate the density of states to be energy
 independent, $g_{\alf\sgm}(\omg)=g_{\alf\sgm}$.
 (Real ferromagnets have a structured density of states,
 which only modifies details of our results but not the
 general physical picture.)
 The asymmetry in the density of states is
 characterized by the degree of spin polarization
 $p_\alf=(g_{\alf\upa}-g_{\alf\dwa})/(g_{\alf\upa}+g_{\alf\dwa})$
 with $-1\leq p_\alf\leq1$,
 in which $p_\alf=0$ denotes a nonmagnetic electrode and
 $p_\alf=\pm1$ corresponds to a half-metallic electrode.

 \begin{figure*}
 \begin{center}
 \includegraphics*[scale=1]{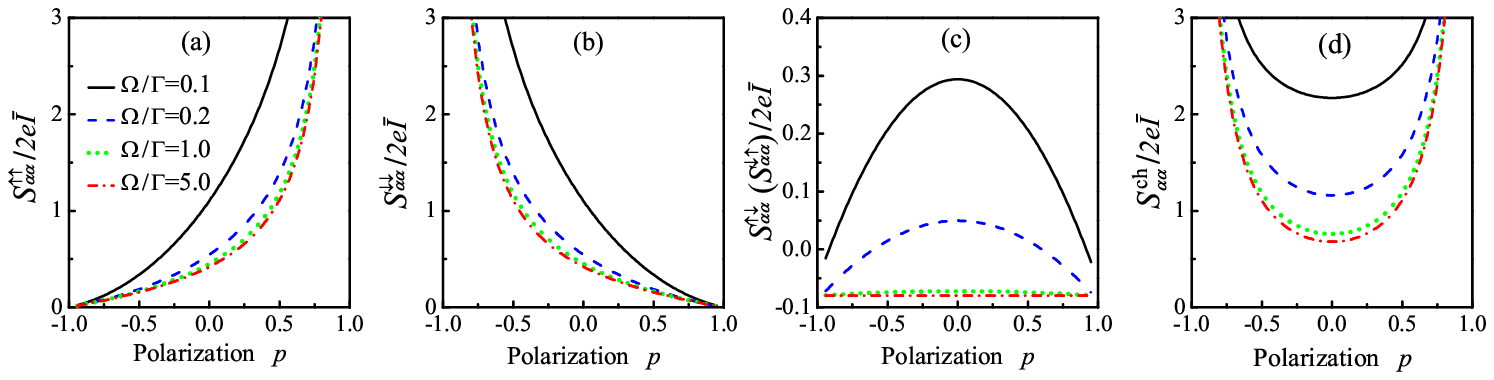}
 \caption{\label{Fig3} Spin self-correlations $S_{\alf\alf}^{\upa\upa}$ and
 $S_{\alf\alf}^{\dwa\dwa}$, spin mutual-correlation $S_{\alf\alf}^{\upa\dwa}$
 ($S_{\alf\alf}^{\dwa\upa}$), as well as
 the total charge current noise $S_{\alf\alf}^{\rm ch}$ vs polarization
 $p$ for various interdot couplings
 $\Omg/\Gam=0.1$, $0.2$, $1.0$, $5.0$.
 The left and right electrodes are of P configuration.
 The other parameters are the same as those in
 \fref{Fig2}.}
 \end{center}
 \end{figure*}

 The Hamiltonian for the coupled dots reads
 \bea
 \hat{H}_{\rm S}&=&\sum_{\ell=1,2}\left[\sum_{\sgm=\upa,\dwa}
 E_\ell \hat{n}_{\ell\sgm}+U_0\hat{n}_{\ell\upa}\hat{n}_{\ell\dwa}\right]
 +U'\hat{n}_1 \hat{n}_2
 \nl
 &&+\Omg\sum_\sgm(d_{1\sgm}^\dag d_{2\sgm}
 +d_{2\sgm}^\dag d_{1\sgm}),
 \eea
 where $d_{\ell\sgm}$ ($d_{\ell\sgm}^\dag$) is the creation
 (annihilation) operator of an electron with spin $\sgm=\upa$
 or $\dwa$ in QD1 ($\ell$=1) or QD2 ($\ell$=2);
 $\hat{n}_{\ell\sgm}=d_{\ell\sgm}^\dag d_{\ell\sgm}$ and
 $\hat{n}_\ell=\sum_\sgm\hat{n}_{\ell\sgm}$ are the
 corresponding particle number operators.
 Each QD has a spin-degenerate level within the bias
 window $V=V_\rmL-V_\rmR$.
 $U_0$ and $U'$  are respectively the intradot and
 interdot Coulomb repulsions; $\Omg$ denotes
 the strength of interdot tunnel-coupling.
 Hereafter, we consider double-dot Coulomb blockade
 regime \cite{Luo07085325,Luo08345215},
 i.e., $U_0$ and $U'$ are large enough such that states
 with two or more electrons in the double dots are not
 allowed.
 The involved states are both dots empty $|0\ra$, one
 electron with spin $\sgm$ in the QD1 $|1\sgm\ra$ or QD2
 $|2\sgm\ra$.
 Experimentally, it can be realized by properly tuning
 the gate and bias voltages \cite{Wie031,Ono021313,Fuj061634,Kop06766}.

 Tunneling between QD1 and electrodes is described by
 $\hat{H}'=\sum_{\alf k\sgm}\big(t_{\alf k\sgm}
 c_{\alf k\sgm}^\dag d_{1\sgm}+{\rm H.c.}\big)$.
 The FM electrodes result in spin-dependent tunnel
 couplings
 $\Gam_{\alf\sgm}=2\pi\sum_k|t_{\alf k\sgm}|^2\delta(\vpl_{\alf k\sgm}-\omg)$.
 In what follows, we consider two magnetic configurations.
 (i) The parallel (P) alignment, when the majority of electrons
 in both electrodes point in the same direction. (ii) The antiparallel
 (AP) case, in which the magnetization of the two electrodes
 are opposite to each other.
 Thus for the P alignment we have
 \bsube
 \be
 \Gam_{\rmL\upa/\rmL\dwa}\!=\!\frac{1}{2}(1 \pm p_\rmL)\GamL
 \;\;{\rm and}\;\;
 \Gam_{\rmR\upa/\rmR\dwa}\!=\!\frac{1}{2}(1 \pm p_\rmR)\GamR,
 \ee
 while for the AP configuration we set
 \be
 \Gam_{\rmL\upa/\rmL\dwa}\!=\!\frac{1}{2}(1 \pm p_\rmL)\GamL
 \;\;{\rm and}\;\;
 \Gam_{\rmR\upa/\rmR\dwa}\!=\!\frac{1}{2}(1 \mp p_\rmR)\GamR.
 \ee
 \esube
 Here $\Gam_\alf=(\Gam_{\alf\upa}+\Gam_{\alf\dwa})$ is
 the total coupling strength regardless the spin orientation.

 \section{\label{thsec3}Results and Discussions}

 In order to get a deep understanding of the spin dynamics and
 fluctuations in transport,
 a Monte Carlo method is employed to simulate the
 real-time single spin tunneling events in the quantum-jump regime.
 We introduce two stochastic point variables
 d$N_{\rmL\sgm}(t)$ and d$N_{\rmR\sgm}(t)$
 (with values either 0 or 1) to stand for, respectively,
 the numbers of spin-$\sgm$ ($\sgm$=$\upa,\dwa$) electron tunneled
 to the double dots from left electrode and that from the double dots
 to the right electrode, during the time interval d$t$.
 The conditional quantum master equation for the
 reduced density matrix reads \cite{Goa01125326}
 \bea
 \rmd \rho_\rmc=&-\rmi {\cal L}\rho_\rmc(t)\rmd t
 -\sum_{\sgm=\upa,\dwa}\{\Gam_{\rmL\sgm}{\cal A}[d^\dag_{1\sgm}]
 +\Gam_{\rmR\sgm}{\cal A}[d_{1\sgm}]
 \nl
 &-{\cal P}_{\rmL\sgm}(t)-{\cal P}_{\rmR\sgm}(t)\}\rho_\rmc(t)\rmd t
 \nl
 &+\sum_{\sgm=\upa,\dwa}\rmd N_{\rmL\sgm}
 \left[\frac{\Gam_{\rmL\sgm}{\cal J}[d^\dag_{1\sgm}]}{{\cal P}_{\rmL\sgm}(t)}-1\right]\rho_{\rmc}(t)
 \nl
 &+\sum_{\sgm=\upa,\dwa}\rmd N_{\rmR\sgm}\left[\frac{\Gam_{\rmR\sgm}{\cal J}[d_{1\sgm}]}{{\cal P}_{\rmR\sgm}(t)}-1\right]\rho_{\rmc}(t),\label{rhoc}
 \eea
 where the attached subscript ``c'' is to indicate that the quantum
 state is conditioned on previous measurement results.
 Here the superoperators are defined as
 ${\cal L}(\cdots)\equiv[\hat{H}_{\rm S},\cdots]$,
 ${\cal J}[X]\rho_{\rmc}\equiv X\rho_{\rmc}X^\dag$ and
 ${\cal A}[X]\rho_{\rmc}\equiv\frac{1}{2}(X^\dag X\rho_{\rmc}+\rho_{\rmc}X^\dag X)$.
 The involving stochastic point variables satisfy
 \bsube
 \bea
 E[\rmd N_{\rmL\sgm}(t)]=&{\cal P}_{\rmL\sgm}(t)\rmd t=\Tr\{ {\cal J}[\sqrt{\Gam_{\rmL\sgm}}d_{1\sgm}^\dag]\rho_{\rmc}\}\rmd t,\label{dNL}
 \\
 E[\rmd N_{\rmR\sgm}(t)]=&{\cal P}_{\rmR\sgm}(t)\rmd t
 =\Tr\{ {\cal J}[\sqrt{\Gam_{\rmR\sgm}}d_{1\sgm}]\rho_{\rmc}\}\rmd t,\label{dNR}
 \eea
 \esube
 where $E[(\cdots)]$ denotes an ensemble average of a large number of
 quantum trajectories.
 In this quantum trajectory approach, spin tunneling events
 condition the future evolution of the system state [see \eref{rhoc}],
 while the instantaneous quantum state conditions
 the measured spin tunneling events through the double dots
 [see \eref{dNL} and \eref{dNR}] in a self-consistent manner.
 One thus is propagating in parallel the information
 of the conditioned (stochastic) state evolution ($\rho_\rmc$)
 and detection record (d$N_{\alf\sgm}$) in
 a single realization of the readout measurement experiment.

 The spin tunneling events (d$N_{\alf\sgm}$) leads
 straightforwardly to the spin-$\sgm$ dependent
 current $I_{\alf}^{\sgm}(t)=e \rmd N_{\alf\sgm}(t)/\rmd t$,
 and consequently the total charge current
 $I^{\rm ch}_\alf=I^{\upa}_\alf+I^{\dwa}_\alf$.
 Hereafter, we will use
 $\bar{I}\equiv E[I^{\rm ch}_\alf(t)]_{t\rightarrow\infty}$
 to represent the ensemble stationary current.
 The spin-resolved time correlation function
 [$C_{\alf\alf'}^{\sgm\sgm'}(t)$] and
 its corresponding noise spectrum
 [$S_{\alf\alf'}^{\sgm\sgm'}\equiv S_{\alf\alf'}^{\sgm\sgm'}(\omg=0)$]
 can be evaluated following \cite{Goa01235307},
 or by using alternatively a spin-resolved quantum master
 equation approach \cite{Gur05205341,Dju06115327}.
 In what follows, noise between different electrodes are not considered
 as it simply satisfies $S_{\rm LR}^{\sgm\sgm'}=-S_{\alf\alf}^{\sgm\sgm'}$
 for the present two-terminal device
 (Note such a relation generally does not hold for
 a multi-terminal structure \cite{Bag03085316,Cot04115315,Cot04206801}).
 For simplicity, we assume symmetric tunnel couplings
 ($\GamL=\GamR=\Gam/2$) and equal magnitude of spin
 polarization in the two electrodes ($p_\rmL=p_\rmR=p$).

 First let us focus on the P alignment.
 \Fref{Fig2}(a)-(f) show the set of measured spin tunneling
 events ($\rmd N_{\rmL\upa/\rmL\dwa}$) and the
 corresponding real-time quantum state ($\rho_\rmc$) for
 polarization $p=0.8$ and interdot coupling $\Omg/\Gam=1.0$.
 We observe unambiguously the bunching of up-spin tunneling
 events.
 When a down spin is injected into the double dots,
 it will stay there and experiences some oscillations
 between the two dots, until it finally escapes to the
 right electrode
 [see \fref{Fig2} (e) and (f) the instantaneous
 quantum states of a down spin in QD1 and QD2].
 The up spins can flow only in short
 time windows where the current is not blockaded by a
 down spin [see \fref{Fig2}(a)-(c)], leading thus to
 the conventional dynamical spin
 blockade, as discussed in Refs. [\onlinecite{Cot04115315,Cot04206801,Don09153305}].
 For clarity, we refer to this mechanism as dynamical
 spin $\upa$-$\dwa$ blockade to specify the bunching
 of up spins due to dynamical blockade of a down spin.
 The dynamical spin $\upa$-$\dwa$ blockade gives rise to
 a prominent super-Poisson spin self-correlation
 $S^{\upa\upa}_{\alf\alf}$, as shown in \fref{Fig3}(a).
 Note here we only need to consider $S^{\upa\upa}_{\alf\alf}$
 due to the fact that $S^{\upa\upa}_{\alf\alf}$
 and $S^{\dwa\dwa}_{\alf\alf}$ are symmetric
 with respect to the spin polarization $p$, i.e.,
 \fref{Fig3}(a) vs (b).

 The spin self-correlation $S^{\upa\upa}_{\alf\alf}$
 increases monotonically with the polarization as
 displayed in \fref{Fig3}(a).
 Only for sufficient spin polarization that the ``$\upa$-$\dwa$''
 competition mechanism takes place, which leads eventually to
 the super-Poisson spin
 self-correlation $S^{\upa\upa}_{\alf\alf}$.
 Yet, it is also instructive to investigate the noise at
 low polarization, for instance $p=$0.
 The total charge current noise $S_{\alf\alf}^{\rm ch}$ as shown
 in \fref{Fig3}(d) exhibits unambiguously super-Poisson statistics
 for $\Omg/\Gam=0.2$ (dashed line).
 It implies bunching of charge tunneling events regardless
 of the spin orientations.
 However, neither of its components ($S^{\upa\upa}_{\alf\alf}$
 or $S^{\dwa\dwa}_{\alf\alf}$) exceeds the Poisson
 value
 [$S^{\upa\upa}_{\alf\alf}/(2e\bar{I})|_{p=0}
 =S^{\dwa\dwa}_{\alf\alf}/(2e\bar{I})|_{p=0}=0.46$].
 It means that the super-Poisson charge noise does not
 stem from the dynamical spin $\upa$-$\dwa$ or
 $\dwa$-$\upa$ bunching.
 In other words, the spin self-correlations $S^{\upa\upa}_{\alf\alf}$
 and $S^{\dwa\dwa}_{\alf\alf}$ do not capture the whole
 picture of spin bunching.

 We ascribe the occurrence of the super-Poisson charge noise
 to a unique dynamical spin $\upa$-$\upa$
 or $\dwa$-$\dwa$ bunching, which is confirmed by the
 numerical simulation of real-time spin tunneling
 as shown in \fref{Fig2}(g)-(j) for $p$=0 and $\Omg/\Gam=0.2$.
 When a down spin is injected into QD1,
 it cycles to the QD2, where it is localized due to weak
 interdot tunnel-coupling. The current is thus blockaded until the
 down spin finally tunnels to the right electrode,
 which is then followed by a bunch of down spins flowing
 through the system, i.e., bunching of \emph{down} spins
 due to dynamical blockade of a \emph{down} spin
 [see \fref{Fig2}(k)-(l)].
 We call this new mechanism the dynamical spin $\dwa$-$\dwa$
 blockade to distinguish it from the $\upa$-$\dwa$ one.
 Similarly, dynamical spin $\upa$-$\upa$ bunching is
 observed as shown in \fref{Fig2}(g)-(i).
 However, the dynamical spin $\upa$-$\upa$ or $\dwa$-$\dwa$
 blockade does not necessarily give rise to super-Poissonian
 spin self-correlations.
 We will reveal this new spin bunching mechanism is
 intimately associated with the spin mutual-correlation
 $S_{\alf\alf}^{\upa\dwa}$,
 which thus can be utilized as an additional diagnostic
 tool to the dynamics and fluctuations in spin transport.

 \begin{figure}
 \begin{center}
 \includegraphics*[scale=0.75]{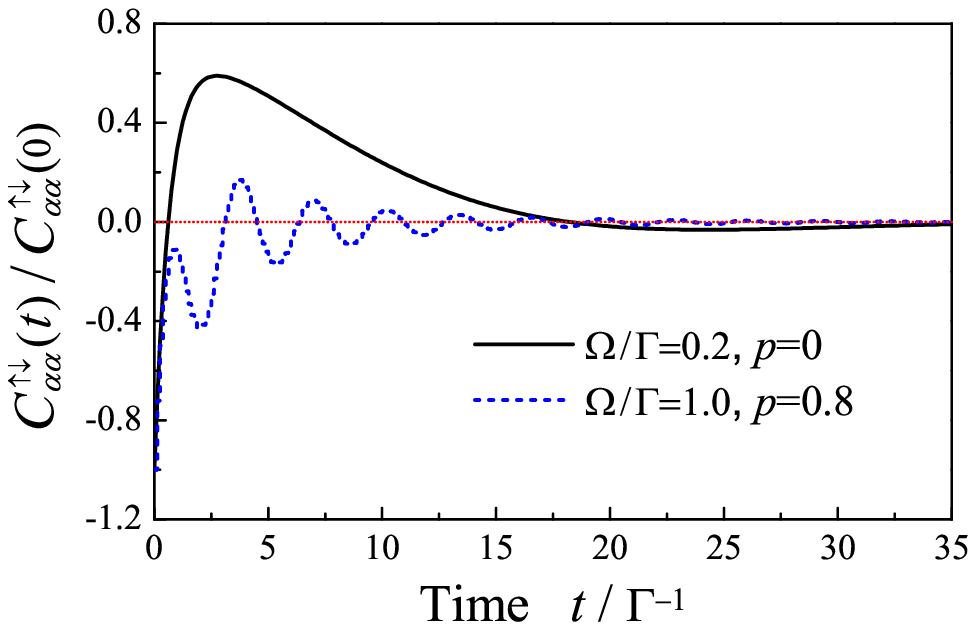}
 \caption{\label{Fig4} Time dependence of correlation function
 $C_{\alf\alf}^{\upa\dwa}(t)$ for $\Omg/\Gam=0.2$, $p=0$ (solid line)
 and $\Omg/\Gam=1.0$, $p=0.8$ (dotted line).}
 \end{center}
 \end{figure}

 For a quantitative analysis, we first evaluate some fundamental
 time scales involved in transport (for $\Omg/\Gam=0.2$ and $p$=0).
 By using 2000 independent trajectories analogous to the
 ones shown in \fref{Fig2}(g)-(l),
 we get the average delay between the occupancy of the
 dots by two consecutive up spins $\tau_0=1.01\Gam^{-1}$
 and the average dwell time of up spins
 on the double dots $\tau_{\upa}=4.00\Gam^{-1}$.
 The average duration of the ``bunch'' of up spins is
 then obtained as $\tau_{\rm b}=6.01\Gam^{-1}$.
 (An alternative approach to obtain these quantities
 can be found in Ref. [\onlinecite{Cot04115315}].)
 The above time scales are able to reveal the
 intrinsic dynamics in spin transport.
 Consider, for instance, the spin-resolved time correlation
 function $C_{\alf\alf}^{\upa\dwa}(t)$, as shown by the
 solid line in \fref{Fig4}.
 It is negative for times shorter than $\tau_0$. It then
 rises, becomes positive and reaches a maximum at
 a time comparable to $\tau_{\upa}$.
 Finally, it decreases on a time scale of
 $\tau_{\upa}+\tau_{\rm b}$.
 For the time scales of tunneling of down spins, analogous
 analysis can be applied.
 The above characteristic times in the correlation function
 thus allow us to attribute the positive $S_{\alf\alf}^{\upa\dwa}$
 to the dynamical spin $\upa$-$\upa$ or $\dwa$-$\dwa$ blockade
 mechanism.
 In comparison, these unique time features can not be identified
 in the case of large interdot coupling ($\Omg/\Gam=1.0$) and
 strong magnetic polarization ($p$=0.8), as shown by the
 dotted line in \fref{Fig4}.

 \begin{figure*}
 \begin{center}
 \includegraphics*[scale=1]{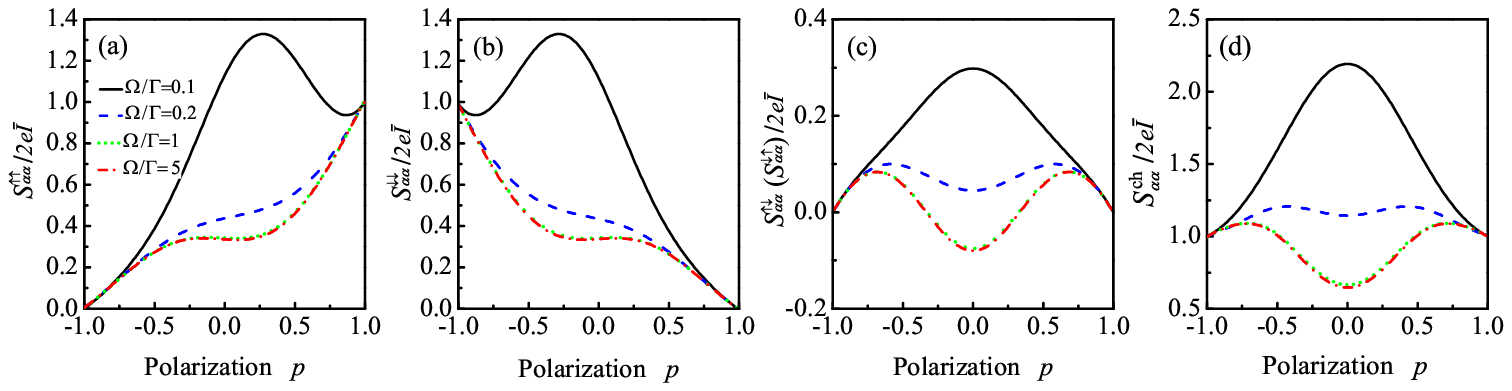}
 \caption{\label{Fig5}
 Spin self-correlations $S_{\alf\alf}^{\upa\upa}$ and $S_{\alf\alf}^{\dwa\dwa}$,
 spin mutual-correlation $S_{\alf\alf}^{\upa\dwa}$ ($S_{\alf\alf}^{\dwa\upa}$),
 as well as the total charge current noise $S_{\alf\alf}^{\rm ch}$
 vs polarization $p$ for various interdot couplings $\Omg/\Gam$
 =0.1, 0.2, 1.0, and 5.0. The left and right electrodes are of AP alignment.
 The other parameters are the same as those in \fref{Fig2}.}
 \end{center}
 \end{figure*}

 The requirements to observe the dynamical spin $\upa$-$\upa$ or
 $\dwa$-$\dwa$ bunching of tunneling events thus can be inferred
 from the spin mutual-correlation. For the P alignment, the
 analytic expression is give by
 \be
 S_{\alf\alf}^{\upa\dwa}=2e\bar{I}\frac{(1-p^2)\Gam^2-16\Omg^2}{200\Omg^2}.
 \ee
 It might be either positive or negative, depending on the
 degree of spin polarization and the strength of interdot
 tunnel-coupling.
 For $\Omg<\frac{1}{4}\Gam$, positive spin mutual-correlation
 is observed provided the electrodes are weakly polarized,
 implying thus the occurrence of dynamical spin $\upa$-$\upa$
 or $\dwa$-$\dwa$ bunching [cf. \fref{Fig2}(g)-(l)].
 It is worth noting that the presented spin transport
 behavior persists over a wide range of polarization
 as long as the interdot coupling is weak enough.
 It thus offers an opportunity to observe the coexistence
 of spin $\upa$-$\upa$ and $\upa$-$\dwa$ bunching of
 tunneling events, as displayed in \fref{Fig2}(m)-(r)
 for spin polarization $p$=0.25 and interdot coupling
 $\Omg/\Gam=0.1$.
 In the opposite regime of $\Omg>\frac{1}{4}\Gam$, the interdot
 coupling is large enough to overcome electron localization
 in QD2, leading thus to the disappearance of the dynamical
 spin $\upa$-$\upa$ or $\dwa$-$\dwa$ bunching, as
 confirmed by our numerical real-time simulation (not
 shown explicitly).
 The resultant spin mutual-correlation is negative at arbitrary
 strength of spin polarization [see the dotted ($\Omg/\Gam=1$)
 and dash-dotted ($\Omg/\Gam=5$) lines in \fref{Fig3}(c)].

 Let us now turn to the situation where the electrodes are
 of AP alignment.
 The self-correlations $S_{\alf\alf}^{\upa\upa}$ and
 $S_{\alf\alf}^{\dwa\dwa}$ vs spin polarization $p$
 are plotted in \fref{Fig5}(a) and (b), respectively.
 Again, we take $S_{\alf\alf}^{\upa\upa}$ for illustration.
 If the left electrode is fully spin-down polarized, transport
 of up spins are completely suppressed, which leads to
 a vanishing $S_{\alf\alf}^{\upa\upa}$ as $p\rightarrow-1$.
 In the opposite limit of $p\rightarrow1$, only up spins are
 allowed to tunnel into the coupled dots; however, under the
 AP alignment the rate of tunneling out to the right electrode
 is strongly inhibited.
 In this case, tunneling of up spins is in close analogy to electron
 tunneling through an extremely asymmetric double barrier
 structure \cite{Che914534}.
 The up spin tunneling events are thus uncorrelated, and the
 resultant noise correlation turns out to be Poisson
 ($S_{\alf\alf}^{\upa\upa}\rightarrow1$), independent of
 interdot coupling strength $\Omg$.
 In a wide range in between, the noise is very sensitive to the
 interdot coupling strength.
 Particularly, we observe a super-Poissonian spin self-correlation
 $S_{\alf\alf}^{\upa\upa}$, as shown by
 the solid line in \fref{Fig5}(a).
 The occurrence of dynamical spin $\upa$-$\dwa$ bunching
 relies on two conditions:
 (i) Appropriate spin polarization in the electrodes, and
 (ii) strong localization of a down spin in QD2, which is
 fulfilled at weak interdot coupling $\Omg$
 ($\Omg<\frac{1}{9}\Gam$ according to our calculation).
 A rising interdot coupling leads to delocalization
 of the down spin. The dynamical spin $\upa$-$\dwa$
 blockade is lifted, which results eventually in
 a sub-Poisson self-correlation for arbitrary
 polarization [see, for instance, the dashed line in
 \fref{Fig5}(a) for $\Omg/\Gam=0.2$].

 Although neither of the two spin self-correlations
 ($S_{\alf\alf}^{\upa\upa}$
 and $S_{\alf\alf}^{\dwa\dwa}$) exhibits super-Poisson statistics
 for $\Omg/\Gam=0.2$,
 the total charge current noise displays unambiguously super-Poisson
 characteristics [dashed line in \fref{Fig5}(d)].
 Analogous to situation of P configuration, the super-Poisson
 noise here arises from the dynamical spin $\upa$-$\upa$ or
 $\dwa$-$\dwa$ bunching, as one can infer from the positive
 spin mutual-correlation [dashed line in \fref{Fig5}(c)].
 A decrease in the strength of interdot tunnel-coupling
 leads to enhancement of the dynamical spin $\upa$-$\upa$
 or $\dwa$-$\dwa$ bunching, and finally increases the spin
 mutual-correlation [cf. \fref{Fig5}].
 Thus, if the interdot tunnel-coupling is low enough one may
 observe the coexistence of dynamical spin $\upa$-$\dwa$
 ($\dwa$-$\upa$) and $\upa$-$\upa$ ($\dwa$-$\dwa$)
 bunching of tunneling events, as indicated by the
 super-Poisson spin self-correlation and positive
 mutual-correlation \fref{Fig5} (a) and (b), respectively.
 Yet, different from the P alignment, the dynamical spin
 $\upa$-$\upa$ blockade survives  even for a strong
 interdot coupling $\Omg$, as long as the
 electrodes are not weakly polarized.
 In this case, an up spin injected into the double dots
 may experience some oscillations between the two dots
 before it can tunnel out to the right electrode owing to
 a small $\Gam_{\rmR\upa}$.
 Its dwell on the double dots serves an ``effective''
 dynamical spin $\upa$-$\upa$ blockade mechanism, yielding
 thus a positive spin mutual-correlation.
 On the other hand, the small $\Gam_{\rmR\upa}$ also
 inhibits tunneling of up spins, which explains the
 suppressed spin mutual-correlation observed in \fref{Fig5}(c).

 We note that current transport through similar
 structure has also been investigated recently
 in \cite{Wey08075305,Wey08045310} and super-Poissonian noise
 was reported.
 There, the existence of a very lower tunneling rate between
 QD2 and the electrodes than that between QD1
 and the electrodes leads to finite occupation of the QD2,
 which gives rise eventually to the super-Poissonian fluctuations.
 Although the mechanisms are different, the finally results
 happen to be consistent qualitatively.
 That is, the total charge current noise is larger
 in the P configuration than in the AP one for a
 large polarization, see \fref{Fig3}(d) and \fref{Fig5}(d).

 \section{\label{thsec4}Conclusion }

 In the context of spin-dependent transport though a Coulomb
 blockaded double quantum dot system, we revealed unambiguously
 unique dynamical spin $\upa$-$\upa$ and $\dwa$-$\dwa$ bunching
 of tunneling events, as confirmed by the
 real-time Monte Carlo simulation of spin tunneling.
 Different from the conventional dynamical spin $\upa$-$\dwa$
 bunching, this new mechanism is found to be intimately
 associated with the spin mutual-correlation
 for both (parallel and antiparallel) magnetic configurations.
 Under conditions of low spin polarization and weak interdot
 tunnel-coupling, we demonstrated the coexistence of the dynamical
 spin $\upa$-$\dwa$ and $\upa$-$\upa$ bunching events.
 Our analysis together with recent suggestions on measurement
 of spin-resolved noise correlations \cite{Sau04106601,For05016601,Cov04184406}
 may shed light on possibilities towards feedback control of
 spin transport through quantum dot systems \cite{Bra10060602,Kie12091226}.

 \begin{acknowledgments}
  This work was supported by the
 National Natural Science Foundation of China
 (grant nos 11204272, 11147114, and 11004124)
 and the Zhejiang Provincial Natural Science Foundation
 (grant nos Y6100171, Y6110467, and LY12A04008).
 \end{acknowledgments}


\begin{thebibliography}{58}
\expandafter\ifx\csname natexlab\endcsname\relax\def\natexlab#1{#1}\fi
\expandafter\ifx\csname bibnamefont\endcsname\relax
  \def\bibnamefont#1{#1}\fi
\expandafter\ifx\csname bibfnamefont\endcsname\relax
  \def\bibfnamefont#1{#1}\fi
\expandafter\ifx\csname citenamefont\endcsname\relax
  \def\citenamefont#1{#1}\fi
\expandafter\ifx\csname url\endcsname\relax
  \def\url#1{\texttt{#1}}\fi
\expandafter\ifx\csname urlprefix\endcsname\relax\def\urlprefix{URL }\fi
\providecommand{\bibinfo}[2]{#2}
\providecommand{\eprint}[2][]{\url{#2}}

\bibitem[{\citenamefont{Blanter and {B\"{u}ttiker}}(2000)}]{Bla001}
\bibinfo{author}{\bibfnamefont{Y.~M.} \bibnamefont{Blanter}} \bibnamefont{and}
  \bibinfo{author}{\bibfnamefont{M.}~\bibnamefont{{B\"{u}ttiker}}},
  \bibinfo{journal}{Phys. Rep.} \textbf{\bibinfo{volume}{336}},
  \bibinfo{pages}{1} (\bibinfo{year}{2000}).

\bibitem[{\citenamefont{Nazarov}(2003)}]{Naz03}
\bibinfo{author}{\bibfnamefont{Y.~V.} \bibnamefont{Nazarov}},
  \emph{\bibinfo{title}{Quantum Noise in Mesoscopic Physics}}
  (\bibinfo{publisher}{Kluwer}, \bibinfo{address}{Dordrecht},
  \bibinfo{year}{2003}).

\bibitem[{\citenamefont{Chen and Ting}(1991)}]{Che914534}
\bibinfo{author}{\bibfnamefont{L.~Y.} \bibnamefont{Chen}} \bibnamefont{and}
  \bibinfo{author}{\bibfnamefont{C.~S.} \bibnamefont{Ting}},
  \bibinfo{journal}{Phys. Rev. B} \textbf{\bibinfo{volume}{43}},
  \bibinfo{pages}{4534} (\bibinfo{year}{1991}).

\bibitem[{\citenamefont{Oliver et~al.}(1999)\citenamefont{Oliver, Kim, Liu, and
  Yamamoto}}]{Oli99299}
\bibinfo{author}{\bibfnamefont{W.~D.} \bibnamefont{Oliver}},
  \bibinfo{author}{\bibfnamefont{J.}~\bibnamefont{Kim}},
  \bibinfo{author}{\bibfnamefont{R.~C.} \bibnamefont{Liu}}, \bibnamefont{and}
  \bibinfo{author}{\bibfnamefont{Y.}~\bibnamefont{Yamamoto}},
  \bibinfo{journal}{Science} \textbf{\bibinfo{volume}{284}},
  \bibinfo{pages}{299} (\bibinfo{year}{1999}).

\bibitem[{\citenamefont{Henny et~al.}(1999)\citenamefont{Henny, Oberholzer,
  Strunk, Heinzel, Ensslin, Holland, and {Sch\"{o}nenberger}}}]{Hen99296}
\bibinfo{author}{\bibfnamefont{M.}~\bibnamefont{Henny}},
  \bibinfo{author}{\bibfnamefont{S.}~\bibnamefont{Oberholzer}},
  \bibinfo{author}{\bibfnamefont{C.}~\bibnamefont{Strunk}},
  \bibinfo{author}{\bibfnamefont{T.}~\bibnamefont{Heinzel}},
  \bibinfo{author}{\bibfnamefont{K.}~\bibnamefont{Ensslin}},
  \bibinfo{author}{\bibfnamefont{M.}~\bibnamefont{Holland}}, \bibnamefont{and}
  \bibinfo{author}{\bibfnamefont{C.}~\bibnamefont{{Sch\"{o}nenberger}}},
  \bibinfo{journal}{Science} \textbf{\bibinfo{volume}{284}},
  \bibinfo{pages}{296} (\bibinfo{year}{1999}).

\bibitem[{\citenamefont{van~der Wiel et~al.}(2003)\citenamefont{van~der Wiel,
  Franceschi, Elzerman, Fujisawa, Tarucha, and Kouwenhoven}}]{Wie031}
\bibinfo{author}{\bibfnamefont{W.~G.} \bibnamefont{van~der Wiel}},
  \bibinfo{author}{\bibfnamefont{S.~D.} \bibnamefont{Franceschi}},
  \bibinfo{author}{\bibfnamefont{J.~M.} \bibnamefont{Elzerman}},
  \bibinfo{author}{\bibfnamefont{T.}~\bibnamefont{Fujisawa}},
  \bibinfo{author}{\bibfnamefont{S.}~\bibnamefont{Tarucha}}, \bibnamefont{and}
  \bibinfo{author}{\bibfnamefont{L.~P.} \bibnamefont{Kouwenhoven}},
  \bibinfo{journal}{Rev. Mod. Phys.} \textbf{\bibinfo{volume}{75}},
  \bibinfo{pages}{1} (\bibinfo{year}{2003}).

\bibitem[{\citenamefont{{Kie{\ss}lich}
  et~al.}(2007)\citenamefont{{Kie{\ss}lich}, {Sch\"{o}ll}, Brandes, Hohls, and
  Haug}}]{Kie07206602}
\bibinfo{author}{\bibfnamefont{G.}~\bibnamefont{{Kie{\ss}lich}}},
  \bibinfo{author}{\bibfnamefont{E.}~\bibnamefont{{Sch\"{o}ll}}},
  \bibinfo{author}{\bibfnamefont{T.}~\bibnamefont{Brandes}},
  \bibinfo{author}{\bibfnamefont{F.}~\bibnamefont{Hohls}}, \bibnamefont{and}
  \bibinfo{author}{\bibfnamefont{R.~J.} \bibnamefont{Haug}},
  \bibinfo{journal}{Phys. Rev. Lett.} \textbf{\bibinfo{volume}{99}},
  \bibinfo{pages}{206602} (\bibinfo{year}{2007}).

\bibitem[{\citenamefont{Lindebaum et~al.}(2009)\citenamefont{Lindebaum, Urban,
  and {K\"{o}nig}}}]{Lin09245303}
\bibinfo{author}{\bibfnamefont{S.}~\bibnamefont{Lindebaum}},
  \bibinfo{author}{\bibfnamefont{D.}~\bibnamefont{Urban}}, \bibnamefont{and}
  \bibinfo{author}{\bibfnamefont{J.}~\bibnamefont{{K\"{o}nig}}},
  \bibinfo{journal}{Phys. Rev. B} \textbf{\bibinfo{volume}{79}},
  \bibinfo{pages}{245303} (\bibinfo{year}{2009}).

\bibitem[{\citenamefont{{Micha{\l}ek} and {Bu{\l}ka}}(2009)}]{Mic09035320}
\bibinfo{author}{\bibfnamefont{G.}~\bibnamefont{{Micha{\l}ek}}}
  \bibnamefont{and} \bibinfo{author}{\bibfnamefont{B.~R.}
  \bibnamefont{{Bu{\l}ka}}}, \bibinfo{journal}{Phys. Rev. B}
  \textbf{\bibinfo{volume}{80}}, \bibinfo{pages}{035320}
  (\bibinfo{year}{2009}).

\bibitem[{\citenamefont{Luo et~al.}(2011{\natexlab{a}})\citenamefont{Luo, Jiao,
  Shen, Cen, He, and Wang}}]{Luo11145301}
\bibinfo{author}{\bibfnamefont{J.~Y.} \bibnamefont{Luo}},
  \bibinfo{author}{\bibfnamefont{H.~J.} \bibnamefont{Jiao}},
  \bibinfo{author}{\bibfnamefont{Y.}~\bibnamefont{Shen}},
  \bibinfo{author}{\bibfnamefont{G.}~\bibnamefont{Cen}},
  \bibinfo{author}{\bibfnamefont{X.-L.} \bibnamefont{He}}, \bibnamefont{and}
  \bibinfo{author}{\bibfnamefont{C.}~\bibnamefont{Wang}}, \bibinfo{journal}{J.
  Phys.: Condens. Matter} \textbf{\bibinfo{volume}{23}},
  \bibinfo{pages}{145301} (\bibinfo{year}{2011}{\natexlab{a}}).

\bibitem[{\citenamefont{Luo et~al.}(2011{\natexlab{b}})\citenamefont{Luo, Shen,
  He, Li, and Yan}}]{Luo1159}
\bibinfo{author}{\bibfnamefont{J.~Y.} \bibnamefont{Luo}},
  \bibinfo{author}{\bibfnamefont{Y.}~\bibnamefont{Shen}},
  \bibinfo{author}{\bibfnamefont{X.-L.} \bibnamefont{He}},
  \bibinfo{author}{\bibfnamefont{X.-Q.} \bibnamefont{Li}}, \bibnamefont{and}
  \bibinfo{author}{\bibfnamefont{Y.~J.} \bibnamefont{Yan}},
  \bibinfo{journal}{Phys. Lett. A} \textbf{\bibinfo{volume}{376}},
  \bibinfo{pages}{59} (\bibinfo{year}{2011}{\natexlab{b}}).

\bibitem[{\citenamefont{Lambert and Nori}(2008)}]{Lam08214302}
\bibinfo{author}{\bibfnamefont{N.}~\bibnamefont{Lambert}} \bibnamefont{and}
  \bibinfo{author}{\bibfnamefont{F.}~\bibnamefont{Nori}},
  \bibinfo{journal}{Phys. Rev. B} \textbf{\bibinfo{volume}{78}},
  \bibinfo{pages}{214302} (\bibinfo{year}{2008}).

\bibitem[{\citenamefont{Prinz}(1998)}]{Pri981660}
\bibinfo{author}{\bibfnamefont{G.~A.} \bibnamefont{Prinz}},
  \bibinfo{journal}{Science} \textbf{\bibinfo{volume}{282}},
  \bibinfo{pages}{1660} (\bibinfo{year}{1998}).

\bibitem[{\citenamefont{Wolf et~al.}(2001)\citenamefont{Wolf, Awschalom,
  Buhrman, Daughton, von {Motn\'{a}r}, Roukes, Chtchelkanova, and
  Treger}}]{Wol011488}
\bibinfo{author}{\bibfnamefont{S.~A.} \bibnamefont{Wolf}},
  \bibinfo{author}{\bibfnamefont{D.~D.} \bibnamefont{Awschalom}},
  \bibinfo{author}{\bibfnamefont{R.~A.} \bibnamefont{Buhrman}},
  \bibinfo{author}{\bibfnamefont{J.~M.} \bibnamefont{Daughton}},
  \bibinfo{author}{\bibfnamefont{S.}~\bibnamefont{von {Motn\'{a}r}}},
  \bibinfo{author}{\bibfnamefont{M.~L.} \bibnamefont{Roukes}},
  \bibinfo{author}{\bibfnamefont{A.~Y.} \bibnamefont{Chtchelkanova}},
  \bibnamefont{and} \bibinfo{author}{\bibfnamefont{D.~M.}
  \bibnamefont{Treger}}, \bibinfo{journal}{Science}
  \textbf{\bibinfo{volume}{294}}, \bibinfo{pages}{1488} (\bibinfo{year}{2001}).

\bibitem[{\citenamefont{Jedema et~al.}(2001)\citenamefont{Jedema, Filip, and
  van Wees}}]{Jed01345}
\bibinfo{author}{\bibfnamefont{F.~J.} \bibnamefont{Jedema}},
  \bibinfo{author}{\bibfnamefont{A.~T.} \bibnamefont{Filip}}, \bibnamefont{and}
  \bibinfo{author}{\bibfnamefont{B.~J.} \bibnamefont{van Wees}},
  \bibinfo{journal}{Nature} \textbf{\bibinfo{volume}{410}},
  \bibinfo{pages}{345} (\bibinfo{year}{2001}).

\bibitem[{\citenamefont{Jedema et~al.}(2002)\citenamefont{Jedema, Heersche,
  Filip, Baselmans, and van Wees}}]{Jed02713}
\bibinfo{author}{\bibfnamefont{F.~J.} \bibnamefont{Jedema}},
  \bibinfo{author}{\bibfnamefont{H.~B.} \bibnamefont{Heersche}},
  \bibinfo{author}{\bibfnamefont{A.~T.} \bibnamefont{Filip}},
  \bibinfo{author}{\bibfnamefont{J.~J.~A.} \bibnamefont{Baselmans}},
  \bibnamefont{and} \bibinfo{author}{\bibfnamefont{B.~J.} \bibnamefont{van
  Wees}}, \bibinfo{journal}{Nature} \textbf{\bibinfo{volume}{416}},
  \bibinfo{pages}{713} (\bibinfo{year}{2002}).

\bibitem[{\citenamefont{Awschalom et~al.}(2002)\citenamefont{Awschalom, Loss,
  and Samarth}}]{Aws02}
\bibinfo{author}{\bibfnamefont{D.~D.} \bibnamefont{Awschalom}},
  \bibinfo{author}{\bibfnamefont{D.}~\bibnamefont{Loss}}, \bibnamefont{and}
  \bibinfo{author}{\bibfnamefont{N.}~\bibnamefont{Samarth}},
  \emph{\bibinfo{title}{Semiconductor spintronics and quantum computation}}
  (\bibinfo{publisher}{Springer}, \bibinfo{address}{Berlin},
  \bibinfo{year}{2002}).

\bibitem[{\citenamefont{Morton et~al.}(2011)\citenamefont{Morton, McCamey,
  Eriksson, and Lyon}}]{Mor11345}
\bibinfo{author}{\bibfnamefont{J.~J.~L.} \bibnamefont{Morton}},
  \bibinfo{author}{\bibfnamefont{D.~R.} \bibnamefont{McCamey}},
  \bibinfo{author}{\bibfnamefont{M.~A.} \bibnamefont{Eriksson}},
  \bibnamefont{and} \bibinfo{author}{\bibfnamefont{S.~A.} \bibnamefont{Lyon}},
  \bibinfo{journal}{Nature} \textbf{\bibinfo{volume}{479}},
  \bibinfo{pages}{345} (\bibinfo{year}{2011}).

\bibitem[{\citenamefont{Hanson et~al.}(2007)\citenamefont{Hanson, Kouwenhoven,
  Petta, Tarucha, and Vandersypen}}]{Han071217}
\bibinfo{author}{\bibfnamefont{R.}~\bibnamefont{Hanson}},
  \bibinfo{author}{\bibfnamefont{L.~P.} \bibnamefont{Kouwenhoven}},
  \bibinfo{author}{\bibfnamefont{J.~R.} \bibnamefont{Petta}},
  \bibinfo{author}{\bibfnamefont{S.}~\bibnamefont{Tarucha}}, \bibnamefont{and}
  \bibinfo{author}{\bibfnamefont{L.~M.~K.} \bibnamefont{Vandersypen}},
  \bibinfo{journal}{Rev. Mod. Phys.} \textbf{\bibinfo{volume}{79}},
  \bibinfo{pages}{1217} (\bibinfo{year}{2007}).

\bibitem[{\citenamefont{\u{Z}uti\'{c} et~al.}(2004)\citenamefont{\u{Z}uti\'{c},
  Fabian, and Sarma}}]{Zut04323}
\bibinfo{author}{\bibfnamefont{I.}~\bibnamefont{\u{Z}uti\'{c}}},
  \bibinfo{author}{\bibfnamefont{J.}~\bibnamefont{Fabian}}, \bibnamefont{and}
  \bibinfo{author}{\bibfnamefont{S.~D.} \bibnamefont{Sarma}},
  \bibinfo{journal}{Rev. Mod. Phys.} \textbf{\bibinfo{volume}{76}},
  \bibinfo{pages}{323} (\bibinfo{year}{2004}).

\bibitem[{\citenamefont{Thielmann et~al.}(2005)\citenamefont{Thielmann,
  Hettler, {K\"{o}nig}, and {Sch\"{o}n}}}]{Thi05146806}
\bibinfo{author}{\bibfnamefont{A.}~\bibnamefont{Thielmann}},
  \bibinfo{author}{\bibfnamefont{M.~H.} \bibnamefont{Hettler}},
  \bibinfo{author}{\bibfnamefont{J.}~\bibnamefont{{K\"{o}nig}}},
  \bibnamefont{and}
  \bibinfo{author}{\bibfnamefont{G.}~\bibnamefont{{Sch\"{o}n}}},
  \bibinfo{journal}{Phys. Rev. Lett.} \textbf{\bibinfo{volume}{95}},
  \bibinfo{pages}{146806} (\bibinfo{year}{2005}).

\bibitem[{\citenamefont{Braun et~al.}(2006)\citenamefont{Braun, {K\"{o}nig},
  and Martinek}}]{Bra06075328}
\bibinfo{author}{\bibfnamefont{M.}~\bibnamefont{Braun}},
  \bibinfo{author}{\bibfnamefont{J.}~\bibnamefont{{K\"{o}nig}}},
  \bibnamefont{and} \bibinfo{author}{\bibfnamefont{J.}~\bibnamefont{Martinek}},
  \bibinfo{journal}{Phys. Rev. B} \textbf{\bibinfo{volume}{74}},
  \bibinfo{pages}{075328} (\bibinfo{year}{2006}).

\bibitem[{\citenamefont{Weymann et~al.}(2012)\citenamefont{Weymann,
  {Barna\'{s}}, and Krompiewski}}]{Wey12205306}
\bibinfo{author}{\bibfnamefont{I.}~\bibnamefont{Weymann}},
  \bibinfo{author}{\bibfnamefont{J.}~\bibnamefont{{Barna\'{s}}}},
  \bibnamefont{and}
  \bibinfo{author}{\bibfnamefont{S.}~\bibnamefont{Krompiewski}},
  \bibinfo{journal}{Phys. Rev. B} \textbf{\bibinfo{volume}{85}},
  \bibinfo{pages}{205306} (\bibinfo{year}{2012}).

\bibitem[{\citenamefont{Yu and Liang}(2005)}]{Yu05075351}
\bibinfo{author}{\bibfnamefont{H.}~\bibnamefont{Yu}} \bibnamefont{and}
  \bibinfo{author}{\bibfnamefont{J.-Q.} \bibnamefont{Liang}},
  \bibinfo{journal}{Phys. Rev. B} \textbf{\bibinfo{volume}{72}},
  \bibinfo{pages}{075351} (\bibinfo{year}{2005}).

\bibitem[{\citenamefont{Misiorny et~al.}(2009)\citenamefont{Misiorny, Weymann,
  and {Barna\'{s}}}}]{Mis09224420}
\bibinfo{author}{\bibfnamefont{M.}~\bibnamefont{Misiorny}},
  \bibinfo{author}{\bibfnamefont{I.}~\bibnamefont{Weymann}}, \bibnamefont{and}
  \bibinfo{author}{\bibfnamefont{J.}~\bibnamefont{{Barna\'{s}}}},
  \bibinfo{journal}{Phys. Rev. B} \textbf{\bibinfo{volume}{79}},
  \bibinfo{pages}{224420} (\bibinfo{year}{2009}).

\bibitem[{\citenamefont{Weymann and {Barna\'{s}}}(2010)}]{Wey10165450}
\bibinfo{author}{\bibfnamefont{I.}~\bibnamefont{Weymann}} \bibnamefont{and}
  \bibinfo{author}{\bibfnamefont{J.}~\bibnamefont{{Barna\'{s}}}},
  \bibinfo{journal}{Phys. Rev. B} \textbf{\bibinfo{volume}{82}},
  \bibinfo{pages}{165450} (\bibinfo{year}{2010}).

\bibitem[{\citenamefont{Weymann et~al.}(2008)\citenamefont{Weymann,
  {Barna\'{s}}, and Krompiewski}}]{Wey08035422}
\bibinfo{author}{\bibfnamefont{I.}~\bibnamefont{Weymann}},
  \bibinfo{author}{\bibfnamefont{J.}~\bibnamefont{{Barna\'{s}}}},
  \bibnamefont{and}
  \bibinfo{author}{\bibfnamefont{S.}~\bibnamefont{Krompiewski}},
  \bibinfo{journal}{Phys. Rev. B} \textbf{\bibinfo{volume}{78}},
  \bibinfo{pages}{035422} (\bibinfo{year}{2008}).

\bibitem[{\citenamefont{{Lipi\'{n}ski} and Krychowski}(2010)}]{Lip10115327}
\bibinfo{author}{\bibfnamefont{S.}~\bibnamefont{{Lipi\'{n}ski}}}
  \bibnamefont{and}
  \bibinfo{author}{\bibfnamefont{D.}~\bibnamefont{Krychowski}},
  \bibinfo{journal}{Phys. Rev. B} \textbf{\bibinfo{volume}{81}},
  \bibinfo{pages}{115327} (\bibinfo{year}{2010}).

\bibitem[{\citenamefont{Wu et~al.}(2007)\citenamefont{Wu, Queipo, Nasibulin,
  Tsuneta, Wang, Kauppinen, and Hakonen}}]{Wu07156803}
\bibinfo{author}{\bibfnamefont{F.}~\bibnamefont{Wu}},
  \bibinfo{author}{\bibfnamefont{P.}~\bibnamefont{Queipo}},
  \bibinfo{author}{\bibfnamefont{A.}~\bibnamefont{Nasibulin}},
  \bibinfo{author}{\bibfnamefont{T.}~\bibnamefont{Tsuneta}},
  \bibinfo{author}{\bibfnamefont{T.~H.} \bibnamefont{Wang}},
  \bibinfo{author}{\bibfnamefont{E.}~\bibnamefont{Kauppinen}},
  \bibnamefont{and} \bibinfo{author}{\bibfnamefont{P.~J.}
  \bibnamefont{Hakonen}}, \bibinfo{journal}{Phys. Rev. Lett.}
  \textbf{\bibinfo{volume}{99}}, \bibinfo{pages}{156803}
  (\bibinfo{year}{2007}).

\bibitem[{\citenamefont{Weymann et~al.}(2011)\citenamefont{Weymann, {Bu{\l}ka},
  and {Barna\'{s}}}}]{Wey11195302}
\bibinfo{author}{\bibfnamefont{I.}~\bibnamefont{Weymann}},
  \bibinfo{author}{\bibfnamefont{B.~R.} \bibnamefont{{Bu{\l}ka}}},
  \bibnamefont{and}
  \bibinfo{author}{\bibfnamefont{J.}~\bibnamefont{{Barna\'{s}}}},
  \bibinfo{journal}{Phys. Rev. B} \textbf{\bibinfo{volume}{83}},
  \bibinfo{pages}{195302} (\bibinfo{year}{2011}).

\bibitem[{\citenamefont{Weymann and {Barna\'{s}}}(2008)}]{Wey08075305}
\bibinfo{author}{\bibfnamefont{I.}~\bibnamefont{Weymann}} \bibnamefont{and}
  \bibinfo{author}{\bibfnamefont{J.}~\bibnamefont{{Barna\'{s}}}},
  \bibinfo{journal}{Phys. Rev. B} \textbf{\bibinfo{volume}{77}},
  \bibinfo{pages}{075305} (\bibinfo{year}{2008}).

\bibitem[{\citenamefont{Cottet et~al.}(2004{\natexlab{a}})\citenamefont{Cottet,
  Belzig, and Bruder}}]{Cot04115315}
\bibinfo{author}{\bibfnamefont{A.}~\bibnamefont{Cottet}},
  \bibinfo{author}{\bibfnamefont{W.}~\bibnamefont{Belzig}}, \bibnamefont{and}
  \bibinfo{author}{\bibfnamefont{C.}~\bibnamefont{Bruder}},
  \bibinfo{journal}{Phys. Rev. B} \textbf{\bibinfo{volume}{70}},
  \bibinfo{pages}{115315} (\bibinfo{year}{2004}{\natexlab{a}}).

\bibitem[{\citenamefont{Cottet et~al.}(2004{\natexlab{b}})\citenamefont{Cottet,
  Belzig, and Bruder}}]{Cot04206801}
\bibinfo{author}{\bibfnamefont{A.}~\bibnamefont{Cottet}},
  \bibinfo{author}{\bibfnamefont{W.}~\bibnamefont{Belzig}}, \bibnamefont{and}
  \bibinfo{author}{\bibfnamefont{C.}~\bibnamefont{Bruder}},
  \bibinfo{journal}{Phys. Rev. Lett.} \textbf{\bibinfo{volume}{92}},
  \bibinfo{pages}{206801} (\bibinfo{year}{2004}{\natexlab{b}}).

\bibitem[{\citenamefont{Dong et~al.}(2009)\citenamefont{Dong, Lei, and
  Horing}}]{Don09153305}
\bibinfo{author}{\bibfnamefont{B.}~\bibnamefont{Dong}},
  \bibinfo{author}{\bibfnamefont{X.~L.} \bibnamefont{Lei}}, \bibnamefont{and}
  \bibinfo{author}{\bibfnamefont{N.~J.~M.} \bibnamefont{Horing}},
  \bibinfo{journal}{Phys. Rev. B} \textbf{\bibinfo{volume}{80}},
  \bibinfo{pages}{153305} (\bibinfo{year}{2009}).

\bibitem[{\citenamefont{Nauen et~al.}(2002)\citenamefont{Nauen, Hapke-Wurst,
  Hohls, Zeitler, Haug, and Pierz}}]{Nau02161303}
\bibinfo{author}{\bibfnamefont{A.}~\bibnamefont{Nauen}},
  \bibinfo{author}{\bibfnamefont{I.}~\bibnamefont{Hapke-Wurst}},
  \bibinfo{author}{\bibfnamefont{F.}~\bibnamefont{Hohls}},
  \bibinfo{author}{\bibfnamefont{U.}~\bibnamefont{Zeitler}},
  \bibinfo{author}{\bibfnamefont{R.~J.} \bibnamefont{Haug}}, \bibnamefont{and}
  \bibinfo{author}{\bibfnamefont{K.}~\bibnamefont{Pierz}},
  \bibinfo{journal}{Phys. Rev. B} \textbf{\bibinfo{volume}{66}},
  \bibinfo{pages}{161303} (\bibinfo{year}{2002}).

\bibitem[{\citenamefont{Safonov et~al.}(2003)\citenamefont{Safonov, Savchenko,
  Bagrets, Jouravlev, Nazarov, Linfield, and Ritchie}}]{Saf03136801}
\bibinfo{author}{\bibfnamefont{S.~S.} \bibnamefont{Safonov}},
  \bibinfo{author}{\bibfnamefont{A.~K.} \bibnamefont{Savchenko}},
  \bibinfo{author}{\bibfnamefont{D.~A.} \bibnamefont{Bagrets}},
  \bibinfo{author}{\bibfnamefont{O.~N.} \bibnamefont{Jouravlev}},
  \bibinfo{author}{\bibfnamefont{Y.~V.} \bibnamefont{Nazarov}},
  \bibinfo{author}{\bibfnamefont{E.~H.} \bibnamefont{Linfield}},
  \bibnamefont{and} \bibinfo{author}{\bibfnamefont{D.~A.}
  \bibnamefont{Ritchie}}, \bibinfo{journal}{Phys. Rev. Lett.}
  \textbf{\bibinfo{volume}{91}}, \bibinfo{pages}{136801}
  (\bibinfo{year}{2003}).

\bibitem[{\citenamefont{Nauen et~al.}(2004)\citenamefont{Nauen, Hohls,
  {K\"{o}nemann}, and Haug}}]{Nau04113316}
\bibinfo{author}{\bibfnamefont{A.}~\bibnamefont{Nauen}},
  \bibinfo{author}{\bibfnamefont{F.}~\bibnamefont{Hohls}},
  \bibinfo{author}{\bibfnamefont{J.}~\bibnamefont{{K\"{o}nemann}}},
  \bibnamefont{and} \bibinfo{author}{\bibfnamefont{R.~J.} \bibnamefont{Haug}},
  \bibinfo{journal}{Phys. Rev. B} \textbf{\bibinfo{volume}{69}},
  \bibinfo{pages}{113316} (\bibinfo{year}{2004}).

\bibitem[{\citenamefont{Sasaki et~al.}(2009)\citenamefont{Sasaki, Tamura,
  Akazaki, and Fujisawa}}]{Sas09266806}
\bibinfo{author}{\bibfnamefont{S.}~\bibnamefont{Sasaki}},
  \bibinfo{author}{\bibfnamefont{H.}~\bibnamefont{Tamura}},
  \bibinfo{author}{\bibfnamefont{T.}~\bibnamefont{Akazaki}}, \bibnamefont{and}
  \bibinfo{author}{\bibfnamefont{T.}~\bibnamefont{Fujisawa}},
  \bibinfo{journal}{Phys. Rev. Lett.} \textbf{\bibinfo{volume}{103}},
  \bibinfo{pages}{266806} (\bibinfo{year}{2009}).

\bibitem[{\citenamefont{Kimand and Hershfield}(2001)}]{Kim01245326}
\bibinfo{author}{\bibfnamefont{T.-S.} \bibnamefont{Kimand}} \bibnamefont{and}
  \bibinfo{author}{\bibfnamefont{S.}~\bibnamefont{Hershfield}},
  \bibinfo{journal}{Phys. Rev. B} \textbf{\bibinfo{volume}{63}},
  \bibinfo{pages}{245326} (\bibinfo{year}{2001}).

\bibitem[{\citenamefont{Cornaglia and Grempel}(2005)}]{Cor05075305}
\bibinfo{author}{\bibfnamefont{P.~S.} \bibnamefont{Cornaglia}}
  \bibnamefont{and} \bibinfo{author}{\bibfnamefont{D.~R.}
  \bibnamefont{Grempel}}, \bibinfo{journal}{Phys. Rev. B}
  \textbf{\bibinfo{volume}{71}}, \bibinfo{pages}{075305}
  (\bibinfo{year}{2005}).

\bibitem[{\citenamefont{Djuric et~al.}(2005)\citenamefont{Djuric, Dong, and
  Cui}}]{Dju05032105}
\bibinfo{author}{\bibfnamefont{I.}~\bibnamefont{Djuric}},
  \bibinfo{author}{\bibfnamefont{B.}~\bibnamefont{Dong}}, \bibnamefont{and}
  \bibinfo{author}{\bibfnamefont{H.~L.} \bibnamefont{Cui}},
  \bibinfo{journal}{Appl. Phys. Lett.} \textbf{\bibinfo{volume}{87}},
  \bibinfo{pages}{032105} (\bibinfo{year}{2005}).

\bibitem[{\citenamefont{Luo et~al.}(2010)\citenamefont{Luo, Wang, He, Li, and
  Yan}}]{Luo10083720}
\bibinfo{author}{\bibfnamefont{J.~Y.} \bibnamefont{Luo}},
  \bibinfo{author}{\bibfnamefont{S.-K.} \bibnamefont{Wang}},
  \bibinfo{author}{\bibfnamefont{X.-L.} \bibnamefont{He}},
  \bibinfo{author}{\bibfnamefont{X.-Q.} \bibnamefont{Li}}, \bibnamefont{and}
  \bibinfo{author}{\bibfnamefont{Y.~J.} \bibnamefont{Yan}},
  \bibinfo{journal}{J. Appl. Phys.} \textbf{\bibinfo{volume}{108}},
  \bibinfo{pages}{083720} (\bibinfo{year}{2010}).

\bibitem[{\citenamefont{Luo et~al.}(2007)\citenamefont{Luo, Li, and
  Yan}}]{Luo07085325}
\bibinfo{author}{\bibfnamefont{J.~Y.} \bibnamefont{Luo}},
  \bibinfo{author}{\bibfnamefont{X.-Q.} \bibnamefont{Li}}, \bibnamefont{and}
  \bibinfo{author}{\bibfnamefont{Y.~J.} \bibnamefont{Yan}},
  \bibinfo{journal}{Phys. Rev. B} \textbf{\bibinfo{volume}{76}},
  \bibinfo{pages}{085325} (\bibinfo{year}{2007}).

\bibitem[{\citenamefont{Luo et~al.}(2008)\citenamefont{Luo, Li, and
  Yan}}]{Luo08345215}
\bibinfo{author}{\bibfnamefont{J.~Y.} \bibnamefont{Luo}},
  \bibinfo{author}{\bibfnamefont{X.-Q.} \bibnamefont{Li}}, \bibnamefont{and}
  \bibinfo{author}{\bibfnamefont{Y.~J.} \bibnamefont{Yan}},
  \bibinfo{journal}{J. Phys.: Cond. Matt.} \textbf{\bibinfo{volume}{20}},
  \bibinfo{pages}{345215} (\bibinfo{year}{2008}).

\bibitem[{\citenamefont{Ono et~al.}(2002)\citenamefont{Ono, Austing, Tokura,
  and Tarucha}}]{Ono021313}
\bibinfo{author}{\bibfnamefont{K.}~\bibnamefont{Ono}},
  \bibinfo{author}{\bibfnamefont{D.~G.} \bibnamefont{Austing}},
  \bibinfo{author}{\bibfnamefont{Y.}~\bibnamefont{Tokura}}, \bibnamefont{and}
  \bibinfo{author}{\bibfnamefont{S.}~\bibnamefont{Tarucha}},
  \bibinfo{journal}{Science} \textbf{\bibinfo{volume}{297}},
  \bibinfo{pages}{1313} (\bibinfo{year}{2002}).

\bibitem[{\citenamefont{Fujisawa et~al.}(2006)\citenamefont{Fujisawa, Hayashi,
  Tomita, and Hirayama}}]{Fuj061634}
\bibinfo{author}{\bibfnamefont{T.}~\bibnamefont{Fujisawa}},
  \bibinfo{author}{\bibfnamefont{T.}~\bibnamefont{Hayashi}},
  \bibinfo{author}{\bibfnamefont{R.}~\bibnamefont{Tomita}}, \bibnamefont{and}
  \bibinfo{author}{\bibfnamefont{Y.}~\bibnamefont{Hirayama}},
  \bibinfo{journal}{Science} \textbf{\bibinfo{volume}{312}},
  \bibinfo{pages}{1634} (\bibinfo{year}{2006}).

\bibitem[{\citenamefont{Koppens et~al.}(2006)\citenamefont{Koppens, Buizert,
  Tielrooij, Vink, Nowack, Meunier, Kouwenhoven, and Vandersypen}}]{Kop06766}
\bibinfo{author}{\bibfnamefont{F.~H.~L.} \bibnamefont{Koppens}},
  \bibinfo{author}{\bibfnamefont{C.}~\bibnamefont{Buizert}},
  \bibinfo{author}{\bibfnamefont{K.~J.} \bibnamefont{Tielrooij}},
  \bibinfo{author}{\bibfnamefont{I.~T.} \bibnamefont{Vink}},
  \bibinfo{author}{\bibfnamefont{K.~C.} \bibnamefont{Nowack}},
  \bibinfo{author}{\bibfnamefont{T.}~\bibnamefont{Meunier}},
  \bibinfo{author}{\bibfnamefont{L.~P.} \bibnamefont{Kouwenhoven}},
  \bibnamefont{and} \bibinfo{author}{\bibfnamefont{L.~M.~K.}
  \bibnamefont{Vandersypen}}, \bibinfo{journal}{Nature}
  \textbf{\bibinfo{volume}{442}}, \bibinfo{pages}{766} (\bibinfo{year}{2006}).

\bibitem[{\citenamefont{Goan et~al.}(2001)\citenamefont{Goan, Milburn, Wiseman,
  and Sun}}]{Goa01125326}
\bibinfo{author}{\bibfnamefont{H.~S.} \bibnamefont{Goan}},
  \bibinfo{author}{\bibfnamefont{G.~J.} \bibnamefont{Milburn}},
  \bibinfo{author}{\bibfnamefont{H.~M.} \bibnamefont{Wiseman}},
  \bibnamefont{and} \bibinfo{author}{\bibfnamefont{H.~B.} \bibnamefont{Sun}},
  \bibinfo{journal}{Phys. Rev. B} \textbf{\bibinfo{volume}{63}},
  \bibinfo{pages}{125326} (\bibinfo{year}{2001}).

\bibitem[{\citenamefont{Goan and Milburn}(2001)}]{Goa01235307}
\bibinfo{author}{\bibfnamefont{H.~S.} \bibnamefont{Goan}} \bibnamefont{and}
  \bibinfo{author}{\bibfnamefont{G.~J.} \bibnamefont{Milburn}},
  \bibinfo{journal}{Phys. Rev. B} \textbf{\bibinfo{volume}{64}},
  \bibinfo{pages}{235307} (\bibinfo{year}{2001}).

\bibitem[{\citenamefont{Gurvitz et~al.}(2005)\citenamefont{Gurvitz, Mozyrsky,
  and Berman}}]{Gur05205341}
\bibinfo{author}{\bibfnamefont{S.~A.} \bibnamefont{Gurvitz}},
  \bibinfo{author}{\bibfnamefont{D.}~\bibnamefont{Mozyrsky}}, \bibnamefont{and}
  \bibinfo{author}{\bibfnamefont{G.~P.} \bibnamefont{Berman}},
  \bibinfo{journal}{Phys. Rev. B} \textbf{\bibinfo{volume}{72}},
  \bibinfo{pages}{205341} (\bibinfo{year}{2005}).

\bibitem[{\citenamefont{Djuric and Search}(2006)}]{Dju06115327}
\bibinfo{author}{\bibfnamefont{I.}~\bibnamefont{Djuric}} \bibnamefont{and}
  \bibinfo{author}{\bibfnamefont{C.~P.} \bibnamefont{Search}},
  \bibinfo{journal}{Phys. Rev. B} \textbf{\bibinfo{volume}{74}},
  \bibinfo{pages}{115327} (\bibinfo{year}{2006}).

\bibitem[{\citenamefont{Bagrets and Nazarov}(2003)}]{Bag03085316}
\bibinfo{author}{\bibfnamefont{D.~A.} \bibnamefont{Bagrets}} \bibnamefont{and}
  \bibinfo{author}{\bibfnamefont{Y.~V.} \bibnamefont{Nazarov}},
  \bibinfo{journal}{Phys. Rev. B} \textbf{\bibinfo{volume}{67}},
  \bibinfo{pages}{085316} (\bibinfo{year}{2003}).

\bibitem[{\citenamefont{Weymann}(2008)}]{Wey08045310}
\bibinfo{author}{\bibfnamefont{I.}~\bibnamefont{Weymann}},
  \bibinfo{journal}{Phys. Rev. B} \textbf{\bibinfo{volume}{78}},
  \bibinfo{pages}{045310} (\bibinfo{year}{2008}).

\bibitem[{\citenamefont{Sauret and Feinberg}(2004)}]{Sau04106601}
\bibinfo{author}{\bibfnamefont{O.}~\bibnamefont{Sauret}} \bibnamefont{and}
  \bibinfo{author}{\bibfnamefont{D.}~\bibnamefont{Feinberg}},
  \bibinfo{journal}{Phys. Rev. Lett.} \textbf{\bibinfo{volume}{92}},
  \bibinfo{pages}{106601} (\bibinfo{year}{2004}).

\bibitem[{\citenamefont{Foros et~al.}(2005)\citenamefont{Foros, Brataas,
  Tserkovnyak, and Bauer}}]{For05016601}
\bibinfo{author}{\bibfnamefont{J.}~\bibnamefont{Foros}},
  \bibinfo{author}{\bibfnamefont{A.}~\bibnamefont{Brataas}},
  \bibinfo{author}{\bibfnamefont{Y.}~\bibnamefont{Tserkovnyak}},
  \bibnamefont{and} \bibinfo{author}{\bibfnamefont{G.~E.} \bibnamefont{Bauer}},
  \bibinfo{journal}{Phys. Rev. Lett.} \textbf{\bibinfo{volume}{95}},
  \bibinfo{pages}{016601} (\bibinfo{year}{2005}).

\bibitem[{\citenamefont{Covington et~al.}(2004)\citenamefont{Covington,
  AlHajDarwish, Ding, Gokemeijer, and Seigler}}]{Cov04184406}
\bibinfo{author}{\bibfnamefont{M.}~\bibnamefont{Covington}},
  \bibinfo{author}{\bibfnamefont{M.}~\bibnamefont{AlHajDarwish}},
  \bibinfo{author}{\bibfnamefont{Y.}~\bibnamefont{Ding}},
  \bibinfo{author}{\bibfnamefont{N.~J.} \bibnamefont{Gokemeijer}},
  \bibnamefont{and} \bibinfo{author}{\bibfnamefont{M.~A.}
  \bibnamefont{Seigler}}, \bibinfo{journal}{Phys. Rev. B}
  \textbf{\bibinfo{volume}{69}}, \bibinfo{pages}{184406}
  (\bibinfo{year}{2004}).

\bibitem[{\citenamefont{Brandes}(2010)}]{Bra10060602}
\bibinfo{author}{\bibfnamefont{T.}~\bibnamefont{Brandes}},
  \bibinfo{journal}{Phys. Rev. Lett.} \textbf{\bibinfo{volume}{105}},
  \bibinfo{pages}{060602} (\bibinfo{year}{2010}).

\bibitem[{\citenamefont{{Kie{\ss}lich}
  et~al.}(2012)\citenamefont{{Kie{\ss}lich}, Emary, Schaller, and
  Brandes}}]{Kie12091226}
\bibinfo{author}{\bibfnamefont{G.}~\bibnamefont{{Kie{\ss}lich}}},
  \bibinfo{author}{\bibfnamefont{C.}~\bibnamefont{Emary}},
  \bibinfo{author}{\bibfnamefont{G.}~\bibnamefont{Schaller}}, \bibnamefont{and}
  \bibinfo{author}{\bibfnamefont{T.}~\bibnamefont{Brandes}},
  \bibinfo{journal}{LANL e-print arXiv:1209.1226}  (\bibinfo{year}{2012}).

\end{thebibliography}

\end{document}